\documentclass[
 print,
 amsmath,amssymb,
 aps,
]{revtex4-2}
\usepackage[T1]{fontenc}
\usepackage{graphicx}
\usepackage{dcolumn}
\usepackage{bm}
\usepackage{amsmath}
\usepackage{lmodern}
\usepackage{mathtools}
\usepackage{siunitx}
\usepackage{multirow}
\usepackage[colorlinks,citecolor=blue,linkcolor=blue,urlcolor=blue]{hyperref}
\usepackage[normalem]{ulem}

\begin{document}

\title{Neural networks based on ultrafast time-delayed effects in exciton-polaritons}
\author{R.~Mirek$^1$}
\author{A.~Opala$^2$}
\author{M.~Furman$^1$}
\author{M.~Kr\'ol$^1$}
\author{K.~Tyszka$^1$}
\author{B.~Seredy\'nski$^1$}
\author{W.~Pacuski$^1$}
\author{J.~Suffczy\'nski$^1$}
\author{J.~Szczytko$^1$}
\author{M.~Matuszewski$^2$}
\author{B.~Pi\k{e}tka$^1$}
\email{barbara.pietka@fuw.edu.pl}

\affiliation{$^1$Institute of Experimental Physics, Faculty of Physics, \\University of Warsaw, ul. Pasteura 5, PL-02-093 Warsaw, Poland}
\affiliation{$^2$The Institute of Physics, Polish Academy of Sciences,\\ Aleja Lotnik\'ow 32/46, PL-02-668 Warsaw, Poland}
	
	\begin{abstract}
		We demonstrate that time-delayed nonlinear effects in exciton-polaritons can be used to construct neural networks where information is coded in optical pulses arriving consecutively on the sample. The highly nonlinear effects are induced by time-dependent interactions with the excitonic reservoir. These nonlinearities allow to create a nonlinear XOR logic gate that can perform operations on the picosecond timescale. An optoelectronic neural network based on the constructed logic gate performs classification of spoken digits with a high accuracy rate.
	\end{abstract}
	\maketitle

	\section{\label{sec:intro}Introduction} 
	The concept of artificial neural networks (ANN) is a widely used approach to machine learning, designed to deal with large amounts of data. ANNs are highly demanded for various applications such as image and speech recognition, pattern detection, optimization, prediction of stock exchange variations, dedicated advertising, autonomous cars, etc. \cite{LeCun_DeepLearning}. However, speed and efficiency of software ANN implementations is limited by the von Neumann architecture of traditional computers, where memory and computing units are physically separated. To circumvent this issue, neuromorphic computing implements neural networks in hardware where access to external memory is not necessary.
	
	In order to achieve high speed and energy efficiency, photonic systems are natural candidates \cite{Psaltis_Photonic,Feldmann_AllOpticalSpikingNetwork, Soljacic_DeepLearning,Lin,Tait_LightTech_2014,antonik2019human,Wetzstein_review,Shastri_review,Xu_TOPS,McMahon_SinglePhoton}. In electronic devices, communication requires charging of  connection wire capacitance for every bit of information, which leads to significant energy dissipation. At very high data rates, this leads to losses which are unacceptable from the practical point of view. This problem is absent in the case of optical links, where information travels at the speed of light and with almost no energy cost. For this reason, electronic wires are being replaced by optical links at smaller and smaller scales, from long-haul optical fibers, to interconnections in data centers, to direct on-chip optical connections~\cite{sun2015single}. It has also stimulated recent interest in optical computing, leading to remarkable achievements including efficient optical vector-matrix multiplication and neural networks~\cite{Brunner_ReinforcementLearning,Zuo_AllOpticalNN}. 
	One of the main practical obstacles in implementing optical data processing is the weakness of the nonlinear response of optical media, or equivalently photon-photon interactions, which is necessary to implement an activation function of a neuron or a transistor. From this point of view, semiconductor exciton-polaritons, where photons and matter excitations (excitons) coexist in a quantum superposition state \cite{Kavokin_book}, are an exceptionally promising candidates~\cite{Matuszewski_EnergyEfficient}. The excitonic component is providing strong interactions required for low-threshold nonlinear operation \cite{Delteil_NatMat, Dreismann_NatMat}. Thanks to the photonic component, they are able to process data at very short timescales and transport data at the speed of light~\cite{Sanvitto_Transistor,Savvidis_TransistorSwitch,Lagoudakis_RTOrganicTransistor}.
	
	Neural networks based on exciton-polaritons have been recently investigated both theoretically \cite{EspinozaOrtega_PRL2015, Xu_PRA2020, Opala_NeuromorphicComputing,Matuszewski_EnergyEfficient} and experimentally \cite{Ballarini_Neuromorphic, Mirek_XOR, Opala_Backpropagation}. Results presented so far, relied on reservoir computing~\cite{Opala_NeuromorphicComputing} and binarized extreme learning machines~\cite{Huang_ExtremeLearningMachines} approaches where most of the network connections are static and only the output layer is trainable. Recently, it was demonstrated that the highly efficient backpropagation training algorithm can be implemented in an exciton-polariton network, leading to an increase of  prediction accuracy~\cite{Opala_Backpropagation}. 
	
	In experimental realizations presented in the literature, input data has been provided either by light pulses arriving at the same time or continuous light beams shaped by spatial light modulators. However, in many practical applications, such as signal processing or speech recognition, input data has the form of a time-coded series. Moreover, dynamical systems are believed to provide solutions to the problems in miniaturization as they allow to decrease the number of spatial nodes \cite{Appeltant_ncomm}. Therefore, it is important that the computing system is able to process data arriving at different instants. 
	
	In this work, we use the dynamics of an exciton-polaritons to construct a binarized neural network~\cite{Bengio_Binarized, Rastegari} where hidden layer neurons are XOR logic gates processing time coded data. We have recently demonstrated that spatial-coded polariton binarized neural networks can be realized~\cite{Mirek_XOR} reaching state-of-the-art accuracy at the MNIST pattern recognition task. Here, we show that the dynamical properties of polaritons can be exploited to use the time domain, instead of the spatial domain, to process input signal. This is enabled by time-delayed interactions successfully realized in many photonic systems \cite{Brunner_NatComm, Feldmann_AllOpticalSpikingNetwork, Hurtado_SciRep} and also recently in polaritons arranged in spatial lattices \cite{Pavlos_timedelay}. In our realization based on polaritons, the exceptionally strong time-delayed nonlinear effects are provided by the interplay between the timescales of polariton condensate lifetime and the decay of the so-called reservoir of uncondensed excitons. As a proof of principle demonstration, we show that our system is able to make highly accurate classification predictions in a spoken digit recognition task.
	
	\section{\label{sec:exp_details}Sample and experimental details}
	The investigated sample was a semiconductor heterostructure based on CdTe microcavity with two distributed Bragg reflectors (DBR) made of quaternary \mbox{(Cd, Zn, Mg)Te} materials. Microcavity contains three pairs of semimagnetic quantum wells of 20\,nm thickness with small concentration (about $0.5\%$) of manganese ions. The structure was grown on a (100)-oriented GaAs substrate by molecular beam epitaxy. Sample was placed in an optical cryostat at liquid helium temperature and excited nonresonantly with pulses of 3\,ps width at frequency of 76\,MHz. The excitation energy was matching the first minimum of the DBRs stopband on the high-energy side. The emission from the sample was collected on the CCD camera or streak camera.

	\begin{figure}[ht!]
		\centering
		\includegraphics[width=.8\linewidth]{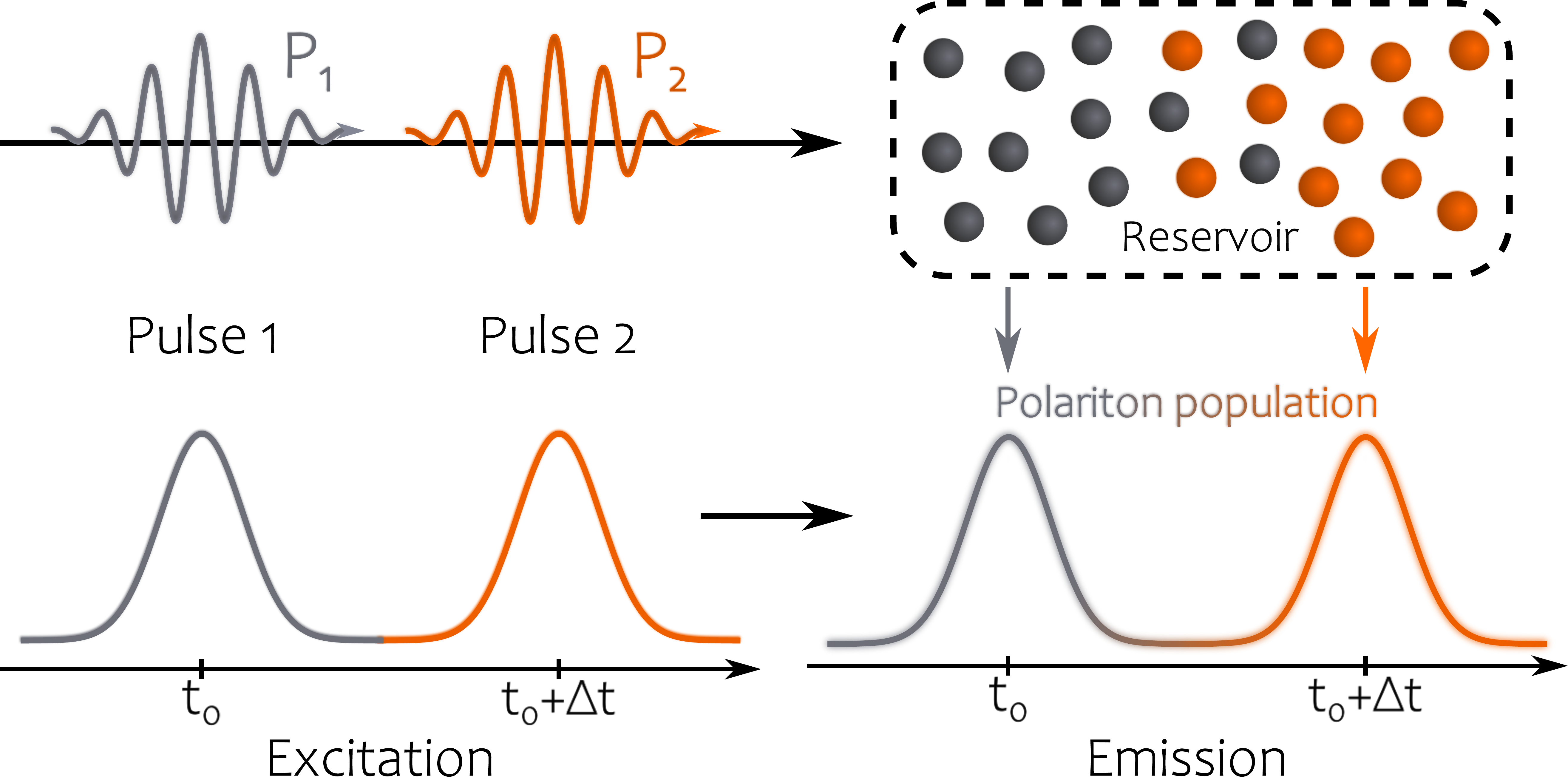}
		\caption{Scheme of the time-delayed interactions. Two different pulses $P_1$ and $P_2$ delayed in time create long-living excitonic reservoir. Relaxation from the reservoir leads to nonlinear polariton population delayed in time.}
		\label{im:scheme}
	\end{figure}

	Our experiment relies directly on the properties of exciton-polaritons. These quasiparticles result from strong coupling of excitons in quantum wells to photon modes in a microcavity~\cite{Kavokin_book}. At high enough densities, they may undergo a transition to a nonequilibrium Bose-Einstein condensate and form a quantum fluid of light~\cite{Carusotto_QuantumFluids}. We create polariton population using two independent laser beams with an adjustable time delay between the pulses, as illustrated in Fig.~\ref{im:scheme}. The first pulse creates long-lived reservoir (uncondensed particles) allowing the formation of exciton-polaritons through energy relaxation and scattering. The reservoir remains populated when the second pulse excites the sample. This leads to the dramatic increase of the density of polaritons and more efficient relaxation to the ground state due to condensation by stimulated scattering. Collective polariton population is therefore coupled through single reservoir, providing time-delayed interactions. 
	
	In the realization of our polaritonic logic gate, the pulses correspond to two inputs where logic states are encoded via excitation power and delay between the pulses. Emission from the condensate serves as the output signal.

	\begin{figure}[t!]
		\centering
		\includegraphics[width=.7\linewidth]{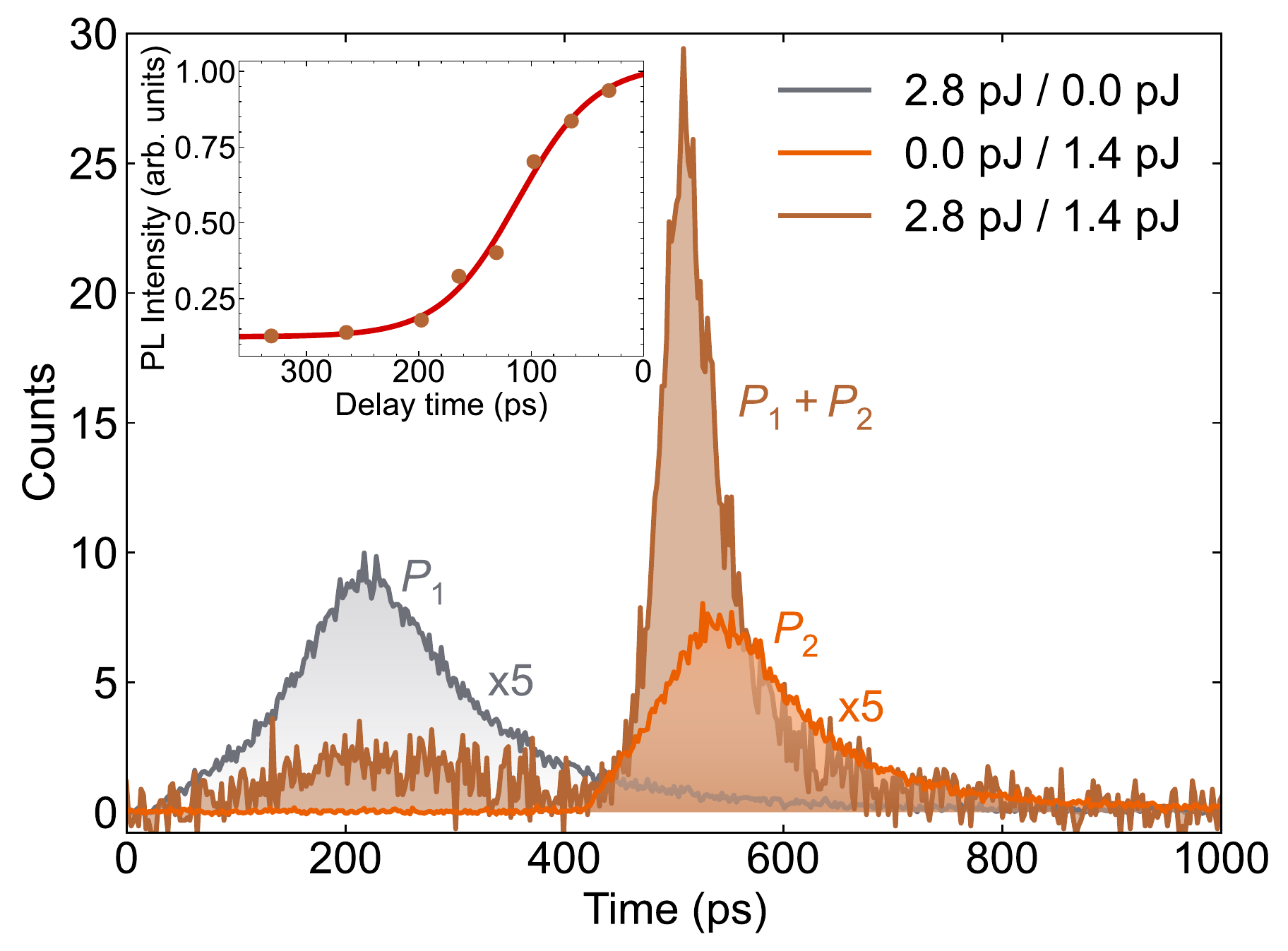}
		\caption{Evidence of nonlinearity. Time-resolved photoluminescence from the exciton-polaritons pumped with the first pulse $P_1$, the second pulse $P_2$ and two time-delayed pulses $P_1+P_2$. The delay between the input pulses was 390.7\,ps. Emission originating from polaritons excited with a single pulse only is increased 5 times for better visibility. The legend shows pulse energy of the first and the second pulse. The inset illustrates intensity of the second emission peak as a function of delay between the pulses.}
		\label{im:NL}
	\end{figure}

	\section{\label{sec:experiment}Time-coded highly nonlinear XOR gate}

	Our method provides large nonlinearities originating from time-delayed interactions of polaritons with long-lived excitonic reservoir and from the build-up of circular polarisation in the condensate. It should be noted that nonlinear effects are negligible below the condensation threshold. Fig.~\ref{im:NL} illustrates the emission signal from exciton-polaritons excited with two pulses delayed in time (brown line). The first peak originates from the weak emission of polaritons below the condensation threshold. The first pulse, however, creates long-lived reservoir that remains active and second pulse increases population of excitons that provide gain to reach condensation threshold. This results in an observation of a second peak with much higher intensity than the first one. To demonstrate the nonlinear nature of this phenomenon, we excite the sample separately with corresponding laser pulses. The gray and orange lines indicate emission originating from excitation with the first and the second pulse separately. Both peaks are multiplied five times for better visibility. We have observed that while the first emission peak has almost the same intensity in both schemes, the second peak induced by a single pulse had significantly lower intensity than the peak resulting from both pulses. The magnitude of nonlinear effects depends on time-delay between the pulses. The inset in Fig.~\ref{im:NL} shows the intensity of the second peak measured for different delays between the pulses. The observed dependence has a strongly nonlinear character and a shape similar to the sigmoid function (marked with red line).
	
	\begin{figure}[t!]
		\centering
        \includegraphics[width=.8\linewidth]{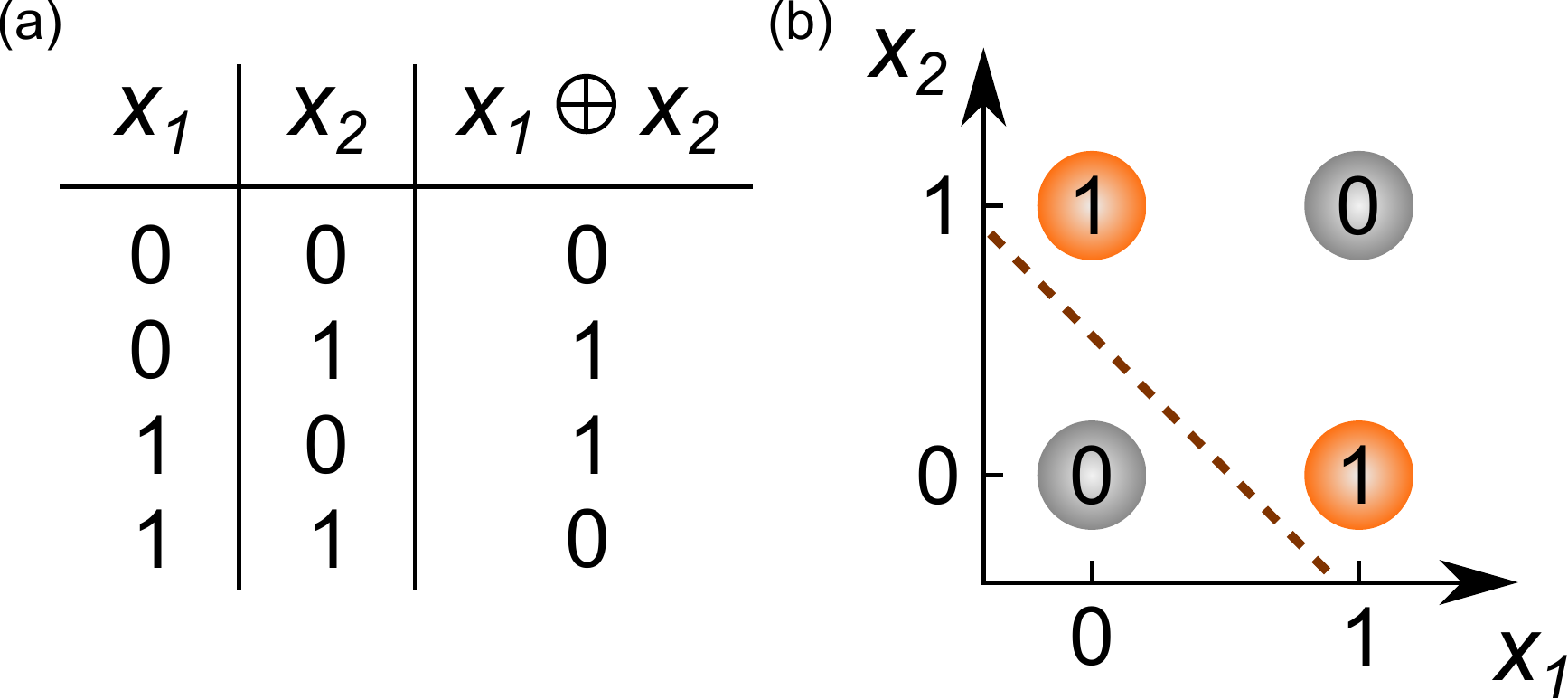}
		\caption{XOR logic gate as a problem not solvable by linear separation. (a) The truth table. (b) The input parameter space and linear separation represented by dashed line.}
		\label{fig:fig_XOR}
	\end{figure}
	\begin{figure}[b!]
		\centering
		\includegraphics[width=.8\linewidth]{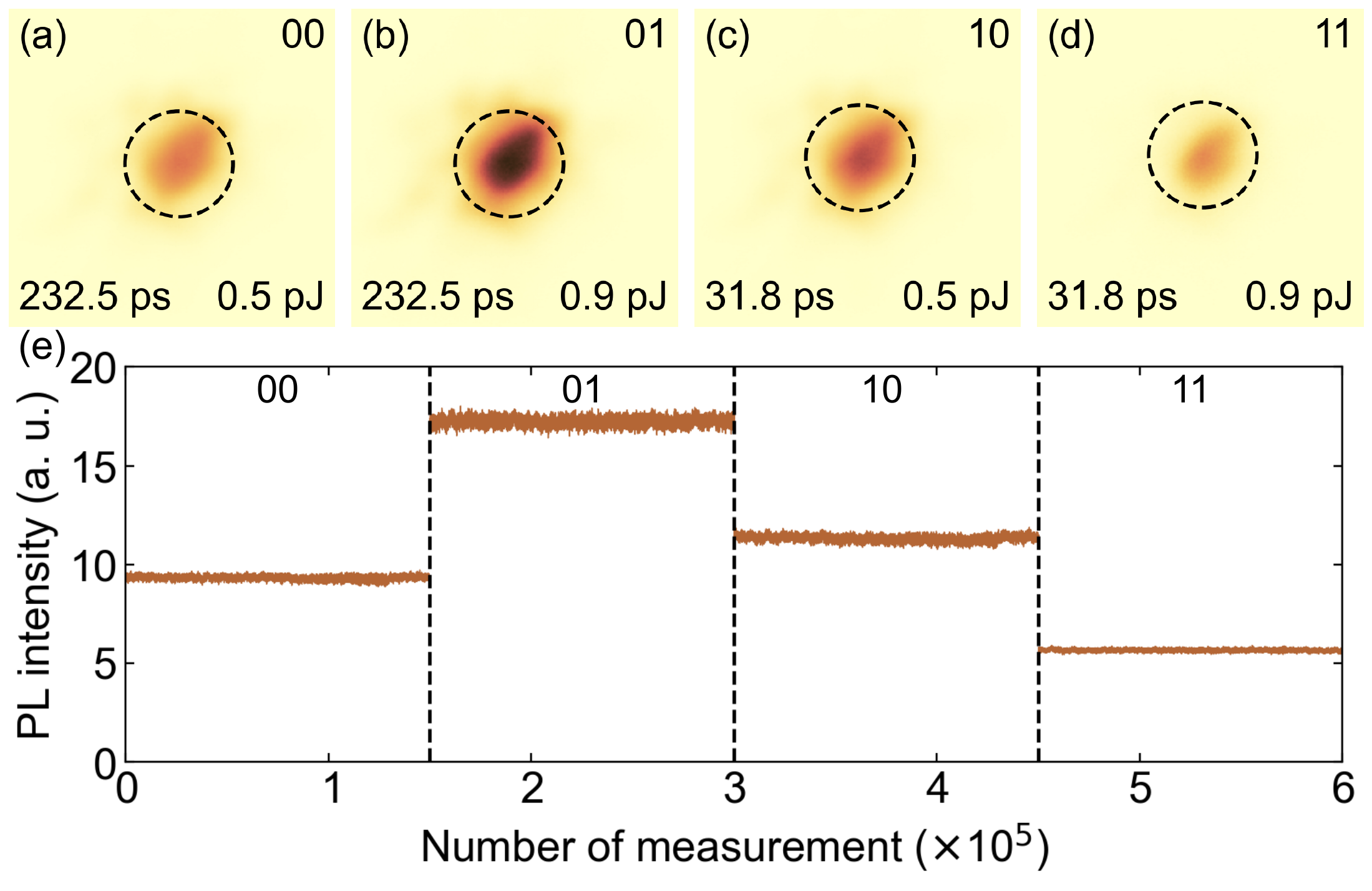}
		\caption{Demonstration of time-coded nonlinearities used for XOR operation. (a)-(d) Real space emission of a polariton condensate with different delay between excitation pulses (indicated at bottom left) and different power of the pulses (bottom right). Corresponding input configurations $x_1x_2$ are marked at top right. Black dashed circles indicate integration regions of 3\,$\mu m$ diameter used for further analysis. Color scale is the same in all panels. (e) Output intensity for all four input combinations.}
		\label{im:XOR}
	\end{figure}
	The nonlinear response of the system is used to construct a XOR gate, which can be considered the simplest form of a nonlinear machine learning problem that cannot be implemented by a simple thresholding of a linear combination of inputs, as is the case for OR~/~AND gates. The XOR classification task is designed to distinguish patterns in two classes, consisting of input combinations ("00", "11") in the first class and ("01", "10") in the second class.  Moreover, it was demonstrated that XOR gates can be used as building blocks of binarized neural networks, which are able to perform complex tasks such as image recognition very efficiently~\cite{Bengio_Binarized,Rastegari}. In this case, XOR gates are used in the role of neurons. The truth table of the XOR problem is shown in Fig.~\ref{fig:fig_XOR}a). The XOR logic gate is an example of a problem that is not linearly separable, as schematically represented in Fig.~\ref{fig:fig_XOR}b). The binary inputs correspond to the $x$ and $y$ axes in the figure. The linear classification task can be performed at most with 75\% accuracy as shown with dashed line. To overcome this problem, we use the nonlinear transformation of inputs by the polariton condensate. A linear combination of gate inputs $x_1,x_2$ and a sufficiently nonlinear output from the condensate $y$ reduce this problem into a linearly separable one in the three-dimensional input-output space~\cite{Mirek_XOR}.
	In this case, the result of the XOR operation is given by the sign of the expression $z=w_1 x_1 + w_2 x_2 + w_3 y(x_1,x_2) + b$, where $w_1$ and $w_2$ are the input weights, $w_3$ is the weight of the nonlinear output feature and $b$ is the bias. 
	The task of finding the proper weights and thresholding are both implemented in software. The weights are determined using a linear regression algorithm. 
	
	The input-output characteristic of a time-coded nonlinear effects for polariton XOR gate is illustrated in Fig.~\ref{im:XOR}. Panels (a)-(d) show real space emission from a spot pumped with $\sigma^+$ polarised pulses and detected in the opposite ($\sigma^-$) circular polarisation. For this polarisation configuration the obtained nonlinearities were the largest. The condensate was excited with two pulsed laser beams with different delay between them (long/short delay encoded 0/1) and two different excitation powers (low/high power encoded 0/1, respectively). Emission was collected on a CCD camera with the acquisition time of 60\,ms. In this case increasing excitation power of laser pulses and decreasing the delay time between them results in a build up of a condensate. Due to the right circular polarisation of input pulses, condensation of semimagnetic exciton-polaritons results in a spontaneous build up of $\sigma^+$ circular polarisation. This leads to a vanishing $\sigma^-$ component as observed in photoluminescence. This agrees with the emission observed in Fig.~\ref{im:XOR}(a)-(d) where the lowest signal intensity was measured for short delay and large power of the pulses. This method allowed us to obtain nonlinearities exceeding significantly noise existing in the system. It is worth noting, that in this case the system dynamics is orders of magnitude faster than the speed of the CCD camera, what is discussed in \ref{sec:Ultrafast}.

	We repeated the experiment to obtain training and testing datasets for the XOR problem. We performed linear classification on a data nonlinearly transformed by polariton condensate. As the nonlinear output signal we used mean intensities in regions marked by black dashed circles in Fig.~\ref{im:XOR}a-d). In Fig.~\ref{im:XOR}e) we show intensities measured in the testing series. Each state was realized 150,000 times in the training phase and 30,000 times in the testing phase. It is important to note, that the output intensities obtained for all input configurations does not have to resemble the XOR gate truth table. At this stage only the high enough nonlinear output is crucial to further perform linear regression and obtain XOR logic gate with perfect accuracy.
	
	The accuracy of the XOR gate depends on the ratio of the obtained nonlinearity to the noise present in the system. We calculated the degree of nonlinearity for a given set of data as~\cite{Mirek_XOR}:
	\begin{equation}
	\eta = \frac{\langle I_{00}\rangle + \langle I_{11}\rangle - \langle I_{01}\rangle - \langle I_{10}\rangle}{\sqrt{V_{00}+V_{01}+V_{10}+V_{11}}},
	\end{equation}
	where $\langle I_{ij}\rangle$ is the average output intensity for the $ij$ input logic state and $V_{ij}$ = $\langle\left( I_{ij} - \langle I_{ij}\rangle\right)^2\rangle$. It was demonstrated previously~\cite{Mirek_XOR} that almost perfect XOR gate operation is achieved for nonlinearity degree  $\eta > 5$. Here during the training phase $\eta\approx 62$. The same procedure was implemented in the testing phase where $\eta\approx57$. The differences between the values in the training and testing sets are originating mostly from the experimental noise. Despite the noise present in the system, it is clear that the proposed system is highly nonlinear. Furthermore, the nonlinearities were large enough to obtain $100\%$ accuracy rate of the XOR gate in the whole testing dataset.
	
	\section{\label{sec:Ultrafast}Ultrafast XOR operation}

\begin{figure}[bt!]
		\centering
		\includegraphics[width=.8\linewidth]{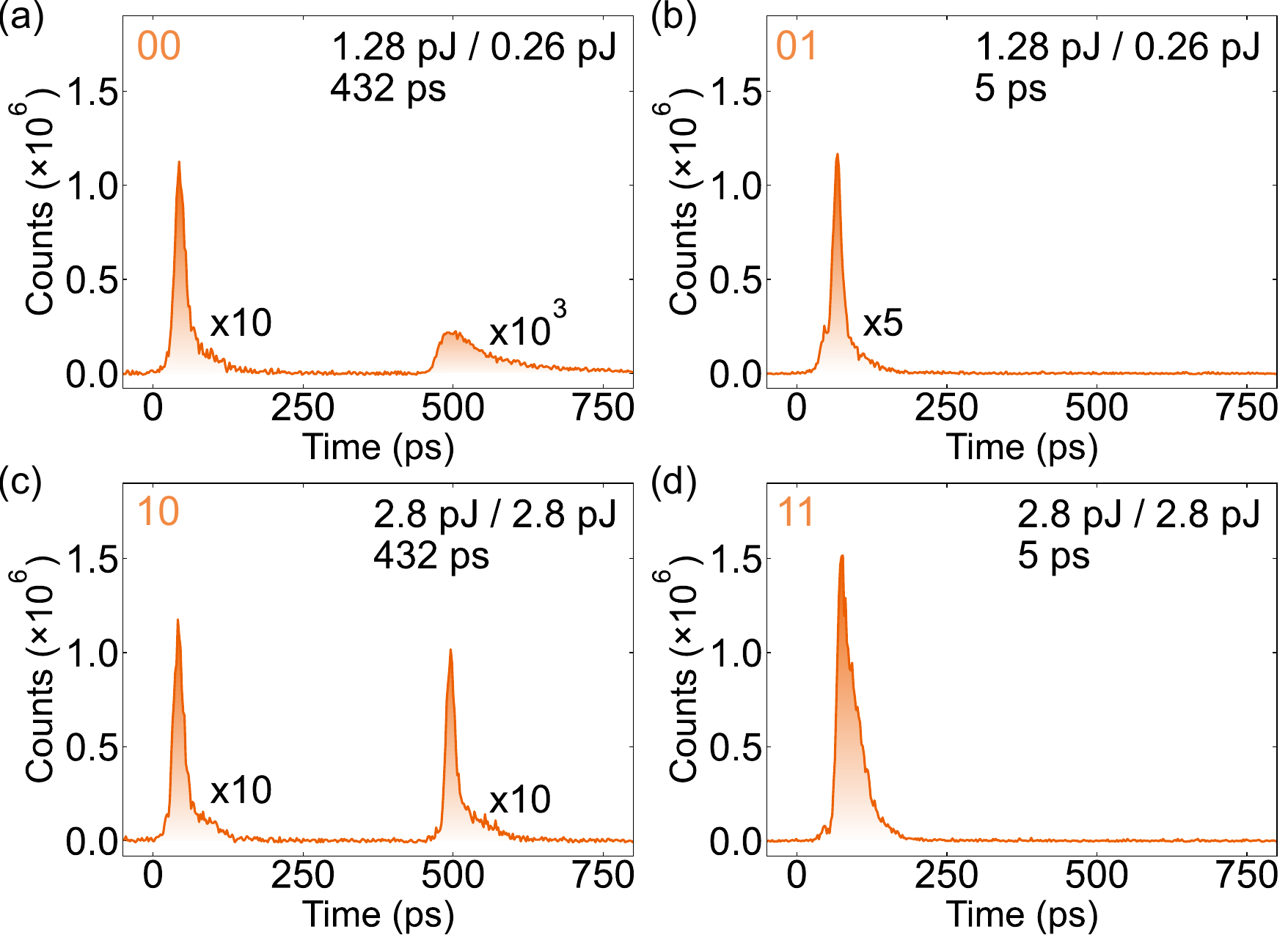}
		\caption{Ultrafast XOR operation. (a)-(d) Time-resolved photoluminescence in the $\sigma^+$ circular polarisation. Excitation power of the two excitation pulses and the delay between them are marked at top right. Top left annotation indicates the realized $x_1x_2$ input configuration. Corresponding peaks are multiplied by marked factors for better visibility.}
		\label{im:streak}
	\end{figure}

	In order to prove the picosecond timescale of the XOR gate operations we performed a time-resolved experiment with a streak camera detection. We studied time-resolved emission in four $x_1x_2$ input configurations coded by different excitation power of the pulses and different delay time between the pulses. Fig.~\ref{im:streak} demonstrates the signal observed in $\sigma^+$ circularly polarised emission. In Fig.~\ref{im:streak}(a), low excitation power and long delay (432\,ps) between the pulses (00) results in the formation of low density exciton-polaritons, which decay over few hundred picoseconds. For two another input configurations (01) and (10), where the delay time was set to 5\,ps or power of the pulses was high, we observed emission with a similar intensities and decay times. On the other hand, the (11) configuration, with short delay between high intensity pulses, resulted in a clearly stronger emission. To calculate the nonlinearites we integrated the signal in the first peak in each configuration which led to $\eta = 70$.
	
	The same experimental conditions were applied during the measurements performed on a spectrometer with a CCD camera. Output was taken as the total intensity observed in the emission. The performed experiments resulted in a degree of nonlinearity above 50, which confirms that the nonlinearities remain high even for the signal averaged over many realizations.
	
	Our results confirm, that the time-coded XOR gate operates in a picosecond scale, but its output can be collected by much slower detector, while still preserving high enough nonlinearities. In fact, there is only one factor playing a crucial role in a XOR gate operation speed. It is a lifetime of an excitonic reservoir which imposes the minimum delay between the laser pulses and for this realization the value is about 250\,ps. In particular, it is possible to create a XOR gate using constant delay between the pulses, defining input parameter space with power of the first and second pulse.
	\section{\label{sec:speech_recognition}Speech recognition}
 
  \begin{figure*}[b!]
    \centering
    \includegraphics[width=1\textwidth]{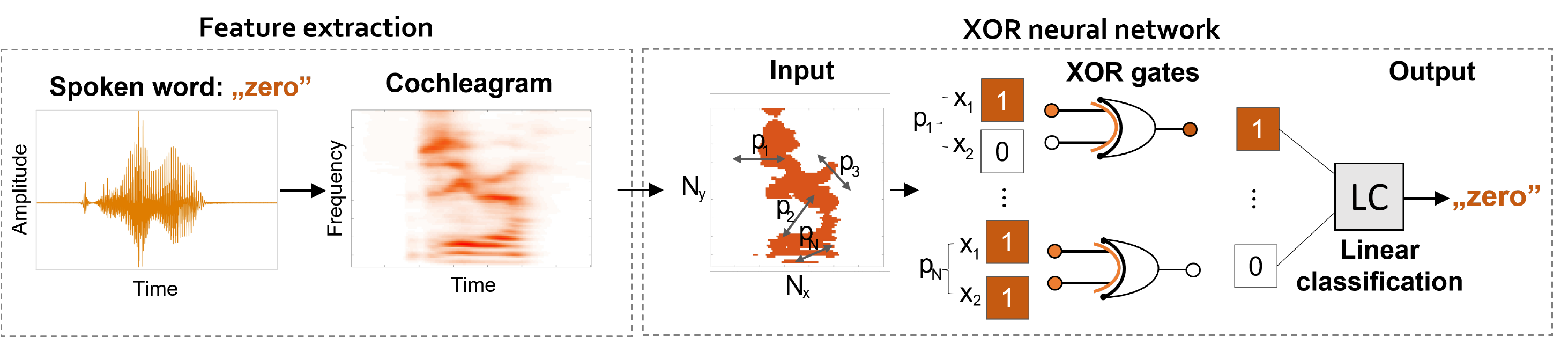}
    \caption{Scheme of the binarized neural network used for the speech recognition task. The recorded audio signal is transformed into a cochleagram and presented as orange and white bitmap. Then the $ N $ randomly selected pixel pairs from the bitmap are used as inputs to a set of XOR logic gates, acting as neurons in the hidden layer. The result of gate operations is used for linear classification in the output layer.}
    \label{fig:fig_1A}
    \end{figure*}
 
    Previously, a binarized exciton-polariton neural network allowed to perform pattern recognition with a high accuracy in comparison to other neuromorphic hardware systems \citep{Mirek_XOR}. This work applies the same type of network architecture, realised using time-multiplexed XOR gates, for speech recognition. To determine the efficiency of our system, we consider a part of the TI 46-Word data set (available from the Linguistic Data Consortium) \citep{TI46}. The database mentioned above is a commonly used speech recognition benchmark for artificial neural networks.
   
    In our analysis, we used 500 training and 1275 testing samples in the form of audio waveform files. Each item is a recorded spoken word that corresponds to one of ten digits (0 to 9). The training set includes words spoken by five different female voices,  each word uttered ten times. The testing set included recordings of 8 other females, repeating digits approximately 16 times.
    
    Using  raw audio files as an input to a neural network is an ineffective method of speech recognition. To obtain optimal classification results, the recorded signals are mapped onto the time-frequency domain using an acoustic transformation \citep{AbreuAraujo2020}. This method can be considered as a feature extraction technique, commonly used in automatic speech recognition. In our work, we employ two methods of audio waveform transformation: the Fourier transform (FT) and the Lyon cochlear-ear model (LCM). The Lyon cochlear-ear model is a nonlinear transformation used to decompose acoustic signals, which mimics the operation of a biological cochlear. It was confirmed that the application of LCM to acoustic signals greatly increases the recognition rate \citep{AbreuAraujo2020}. 
    
    The conceptual scheme of the proposed neural network is shown in Fig.~\ref{fig:fig_1A}. Each audio sample from a data set is pre-processed and transformed into the time-frequency domain using $ N_f $ frequency channels. Next, all maps are normalised and resized to the size $N_f \times N_f$. Time-frequency patterns are converted into eight black and white bitmaps denoted by $ c_i $, where $i= \{1,2, \dots, 8 \}$, and $c_i$ is given by the formula $c_i=\Big| \left \lfloor{\frac{I}{2^{(i-1)}}}\right \rfloor\Big|_2$, where $|a|_b$ is the modulo operation returning the remainder of division of $a/b$. The binary time-frequency maps are used as neural network inputs. From each map we selected $N$ pairs of pixels denoted by $p_1,p_2,p_3,\dots,p_N$ which allow us to reveal nontrivial correlations between pixels in the bitmaps. The positions of the selected pixels did not change during the entire learning process. Next, we use a linear classification, denoted in the figure as LC, to predict the spoken word. To avoid overfitting, we use a stochastic gradient descent method with regularization. As we only train the weights in the output layer used for linear classification, the network is an instance of an extreme learning machine~\cite{Huang_ExtremeLearningMachines}.
    
    \begin{table}[bht!]
    \caption{Classification accuracy determined for spectrograms and cochleagrams using $5000$ XOR gates in the hidden layer. Each value is an average of five different realizations of neural network teaching. Each realization corresponds to a different selection of the randomly selected pixel pairs.}
    \begin{tabular}{|cl|lll|}
    \hline
    \multicolumn{2}{|c|}{Size of time-frequency  map }                               & $(16\times 16)$ & $(32\times 32)$ & $(64 \times 64)$ \\ \hline
    \multicolumn{1}{|c|}{\multirow{2}{*}{Method}} & FT $(\%)$  &  ~~~76.08  &  ~~~87.16  & ~~~94.06   \\ \cline{2-5} 
    \multicolumn{1}{|c|}{}                         & LCM $(\%)$ &  ~~~90.96  &  ~~~95.50  &  ~~~96.44 \\ \hline
    \end{tabular}
    
    \label{tab:tab_1A}
    \end{table}
    
    We examined the efficiency of the network using both spectrograms and cochleagrams as input data. We considered time-frequency maps of size $ (16 \times 16) $, $ (32 \times 32) $ and $ (64 \times 64) $. Data samples delivered to the network were combinations of eight binary maps merged into vectors $ [c_1, c_2, c_3, \dots, c_8] $. The obtained results are presented in Table~\ref{tab:tab_1A}. Increasing the number of XOR gates in the hidden layer results in improvement of accuracy. Accuracy of the spoken words prediction in the case of binarized and not transformed by XOR gates, FT and LCM inputs were: 75.50$\%$, 86.12$\%$, 92.72$\%$ and  86.12$\%$,  94.4$\%$,  95.5$\%$, respectively for $(16\times 16)$, $(32\times 32)$ and $(64\times 64)$ time-frequency bitmaps. The highest increase of correct predictions, equal to 4.84$\%$, was achieved for input in the form of a cochleagram containing 16 frequency channels.
    
	\section{\label{sec:summary}Summary}
	We developed a method for the realization of exciton-polariton based binarized neurons using time-delayed effects. We used time-coded addressing of a single exciton-polariton condensate with laser pulses delayed in time. The system was used to implement two machine learning tasks, the nonlinear XOR problem and spoken digit recognition. Our proof-of-principle experiment demonstrates the capability of exciton-polariton systems to process time-dependent data on the picosecond timescale. 
	
	\section{Acknowledgements}
    This work was supported by National Science Center, Poland:  
    R.M. acknowledges 2019/33/N/ST3/02019, 
    B.P. acknowledges 2017/27/B/ST3/00271, 
    M.K. acknowledges 2015/18/E/ST3/00558, 
    M.F. acknowledges 2020/37/B/ST3/01657, 
    K.T. acknowledges 2020/04/X/ST7/01379,
    A.O. acknowledge 2019/35/N/ST3/01379,
    M.M. acknowledges support from grant no. 2017/25/Z/ST3/03032, under the QuantERA program.
	
	\bibliography{bibliography}

\end{document}